\documentclass[proc]{revtex4}
\usepackage{graphicx}
\usepackage{fancyhdr}
\usepackage{graphics}
\usepackage{epstopdf}
\usepackage{textpos}

\usepackage{xspace}%

\newcommand{\chiOneZero}{\ensuremath{\tilde{\chi}^{0}_{1}}\xspace}

\newcommand{\dxy}{\ensuremath{|d_\textrm{XY}|}~}
\newcommand{\CLs}{\ensuremath{\text{CL}_\text{s}}~}
\newcommand{\PWpr}{\ensuremath{\text{W}'}}%

\newcommand{\Wo}{\PW\xspace}%
\newcommand{\Ho}{\PH\xspace}%
\newcommand{\mjj}{\ensuremath{m_\mathrm{jj}}\xspace}%
\newcommand{\mWH}{\ensuremath{m_{\Wo\Ho}}\xspace}%

\def\mjj{\ensuremath{m_{\text{jj}}}\xspace}

\def\Qstar{\ensuremath{\mbox{q}^*}\xspace}

\input{pdefs}

\begin{document}

\title{\centering Exotic Physics}

\author{
\centering
\begin{center}
Michael Sigamani for the CMS Collaboration
\end{center}}
\affiliation{\centering University of Ghent, Belgium}
\begin{abstract}

A selection of results for searches for exotic physics at the LHC are presented.
These include a search for massive resonances, dark matter with a high energy jet in association with large missing transverse momentum,
long-lived neutral particles, and narrow dijet resonances. 
The results are based on 20 fb$^{-1}$ of LHC proton-proton collisions at $\sqrt{s}=8~$\TeV\
taken with the CMS detector.

\end{abstract}

\maketitle
\thispagestyle{fancy}

\section{Introduction}

The standard model (SM) has been extremely successful at describing particle physics phenomena over the last half-century, 
and the recently discovered boson with a mass of 125~\GeV \cite{CMSHiggs,ATLASHiggs} could be the 
final particle required in this theory, the Higgs boson. 
However, the SM is not without its shortcomings, for instance one requires fine-tuned cancellations of large quantum corrections
in order for the Higgs boson to have a mass at the electroweak
symmetry breaking scale~\cite{SUSY1,SUSY5,SUSY6}. 
This is otherwise known as the hierarchy problem. 
Due to the magnitude of this fine-tuning, one suspects that there is some dynamical mechanism which makes this fine-tuning ``natural''.
Supersymmetry (SUSY) is a popular extension of the SM which postulates
the existence of a sparticle for every SM particle. 
These sparticles have the same quantum numbers as their SM counterparts but differ by one half-unit of spin. 
The loop corrections to the Higgs boson mass due to these sparticles 
are opposite to those of the SM particles thus providing a
natural solution to the hierarchy problem. 
In addition, in $R$-parity~\cite{Farrar:1978xj} conserving SUSY models, the lightest super-symmetric particle (LSP) is often the lightest neutralino \chiOneZero. 
The \chiOneZero offers itself as a good dark matter candidate subsequently 
explaining particular astrophysical observations~\cite{DarkMatterReview,DMGeneral}. 
In this note, several CMS~\cite{Chatrchyan:2008aa} searches are reported for beyond the SM (BSM) physics 
using around 20 fb$^{-1}$ of LHC proton-proton collisions recorded at $\sqrt{s}=8~$\TeV.

\section{Search for long-lived, neutral particles decaying to photons}

The existence of new, heavy particles with long lifetimes is predicted by many BSM theories,
such as hidden valley scenarios~\cite{Strassler:2006im} or gauge-mediated
SUSY breaking (GMSB)~\cite{Giudice:1998bp}.
Under the assumption of R-parity conservation,
strongly interacting SUSY particles can be pair-produced at the Large Hadron Collider (LHC).
Their subsequent decay chain may include one or more quarks and gluons, as well as
the LSP, which escapes detection,
giving rise to apparent momentum imbalance in the transverse plane (\ETm).
A GMSB benchmark scenario, commonly described as ``Snowmass Points and Slopes 8''
(SPS8)~\cite{Allanach:2002nj} is used as reference in this search.
In this scenario, the lightest neutralino (\PSGczDo) is the next-to-lightest SUSY
particle, and can be long-lived. It decays to a photon (or Z/Higgs bosons) and a gravitino (\PXXSG), which is
the LSP~\cite{Dimopoulos:1996vz}. If the \PSGczDo consists predominantly of the bino,
the superpartner of the $U(1)$ gauge field,
its branching fraction to a photon and \PXXSG is expected to be large.

\begin{figure}[!htb]
\centering
\includegraphics[height=0.5\textwidth]{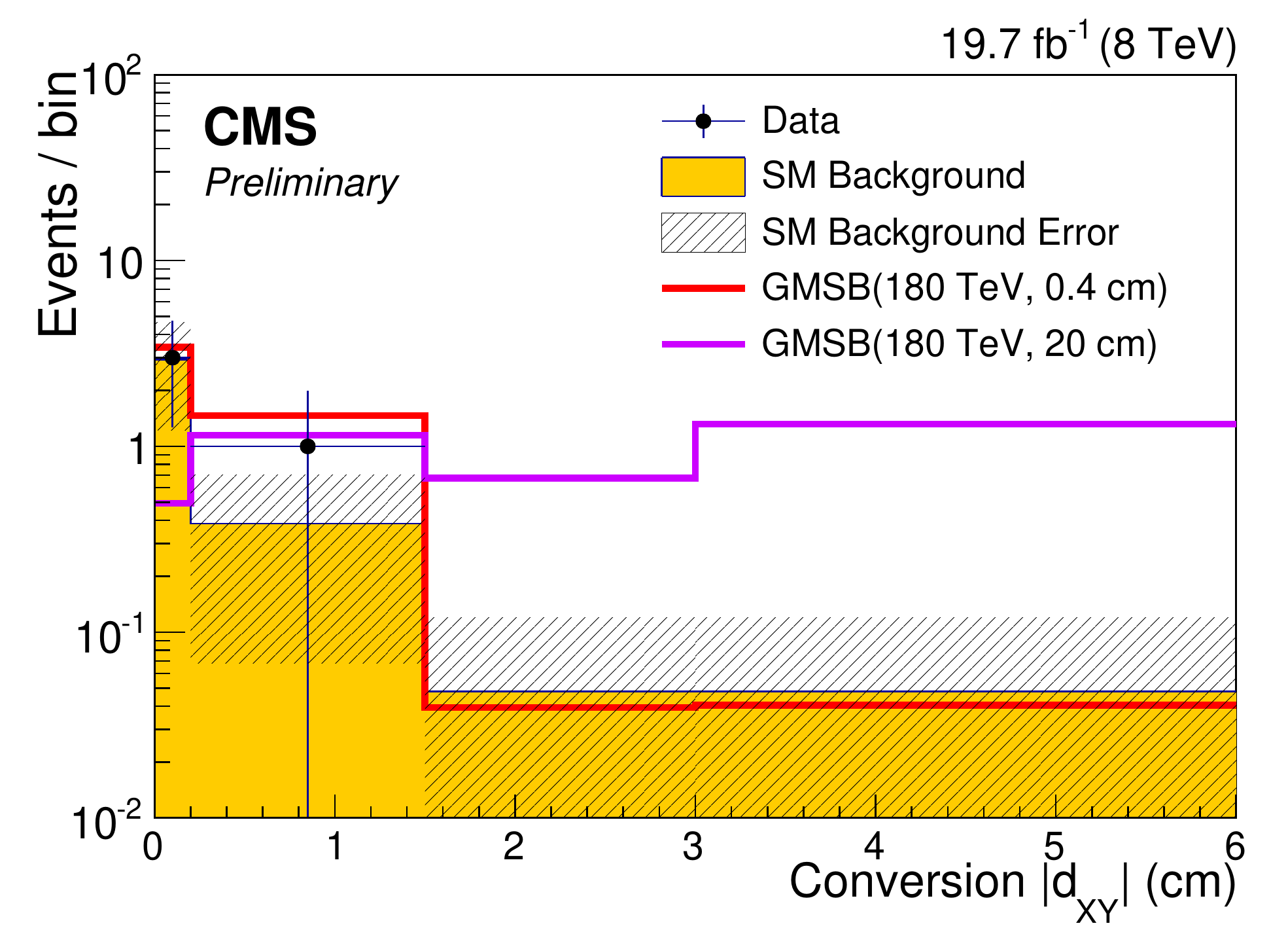}
\caption{Distribution of \dxy: data are compared with background which is estimated from CR1.
   Two signal points for $\Lambda = 180$ TeV, of $c\tau = 0.4$ cm and 20 cm are also shown for comparison.
   The black hatches correspond to the total uncertainty on the
   background estimate. The last bin for each histogram contains the overflow.}
\label{fig:dxyDD}
\end{figure}

A photon produced in the decay of a long-lived \PSGczDo will originate from a displaced vertex
with a large absolute transverse impact parameter, \dxy. In this analysis, reconstructed photon conversions
are exploited to determine the photon direction with respect to the beam axis and to measure the \dxy.
The larger the \PSGczDo lifetime, the wider the distribution of the \dxy.
The strategy consists of selecting events with at least two photons of which at least one has converted to an $e^{+}e^{-}$ pair.
The \dxy distribution is then used to search for an excess of signal over SM background events.
The signature of these decays is extremely clean since it is given by displaced decay vertices
and significant \ETm. The number of signal and background events in data are estimated from the binned \dxy
distributions and shown in Fig.~\ref{fig:dxyDD}. 
The last bin contains also overflows.

\begin{figure}[h]
\centering
\includegraphics[height=0.55\textwidth]{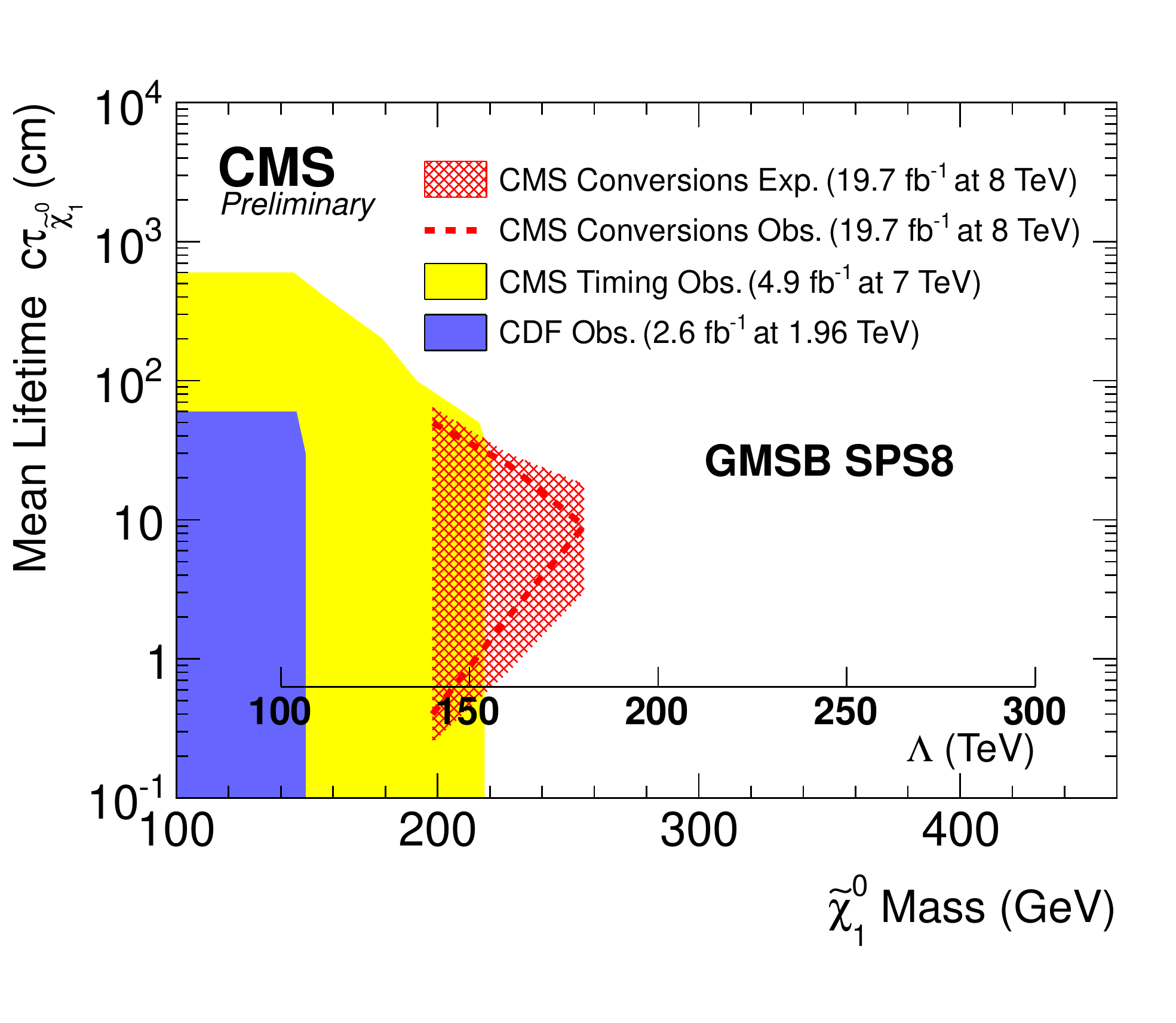}
\caption{Exclusion plot in the plane defined by the $\PSGczDo$ and its mean lifetime in the context of the
  SPS8 model of GMSB supersymmetry. The $\Lambda$ scale is also shown.}
    \label{fig:2dlimit}
\end{figure}

There is no significant excess of events in data with respect to the background expectations, therefore
we interpret the results of the search in terms of
upper limits at the 95\% confidence
level (CL) on the production cross section of a long-lived neutralino
in the context of the GMSB model.
A test of the background-only and signal+background hypotheses is performed using a modified frequentist
approach, often referred to as \CLs~\cite{Junk:1999kv,0954-3899-28-10-313}.
The probability distributions of the background-only and the signal+background
hypotheses are determined from distributions of the test statistic constructed
from pseudo-experiments. Once the ensembles of
pseudo-experiments for the two hypotheses are generated, the observed \CLs
limit is calculated from these distributions and the actual observation
of the test statistic. The expected \CLs limit is calculated
by replacing the test statistic by the expected median from the distribution of the
background-only hypothesis.
Furthermore, since we normalise our background to the
first bin of data, we exclude the signal contribution from this bin in the limit setting procedure.
The uncertainties are handled by introducing nuisance parameters both on the normalisation
and the shape of the signal and background expectations.
The results can be seen in Fig.~\ref{fig:2dlimit}.
Here we plot the region of GMSB phase space where the observed and expected median \CLs limit falls below the theoretical
expectation, and we are therefore able to exclude.
The full study can be found in Ref.~\cite{CMS:2015gga}.

\section{Search for massive WH resonances decaying to the $\ell \nu {\rm b \bar{b}}$ final state in a boosted regime} 

In this section a search for a massive resonance decaying into a W boson and a Higgs boson in the $\ell \nu \rm{b \bar{b}}$ final state is presented. 
Such processes are prominent features of several extensions of the SM
such as Composite Higgs ~\cite{Composite0,Composite1,Composite2} and Little Higgs
models~\cite{Han:2003wu}.
These models provide a direct solution to the hierarchy problem, and
predict many new particles including additional gauge bosons such as a heavy $\PWpr$ boson.
The $\PWpr$ boson in these models can have large decay branching ratios to WH and WZ,
while the decay to fermions is suppressed.
The recently proposed Heavy Vector Triplet (HVT) model~\cite{Pappadopulo:2014qza}
generalizes a large class of explicit models predicting new heavy spin-1 vector bosons.
The resonance is described by a simplified Lagrangian in terms of few relevant parameters
representing its mass and couplings to SM bosons and fermions.
For a $\PWpr$ with the SM couplings to fermions and thus reduced couplings to SM bosons,
the most stringent limits are reported in searches with
leptonic final states~\cite{CMSwprimePAPER2013,ATLASwprimePAPER2014}, and the current lower
limit on the $\PWpr$ mass is 3.2~TeV.
The signal studied in this analysis is the production of a resonance with mass above 0.8~TeV,
decaying to WH where the Higgs boson decays to two bottom quarks and the W boson decays to a lepton and
a neutrino. No assumptions on additional particles produced
in the final state are made. It is assumed that the resonance is narrow, i.e. that its natural
width is much smaller than the experimental resolution.

\begin{figure}[htbp]
\centering
\includegraphics[width=0.45\linewidth]{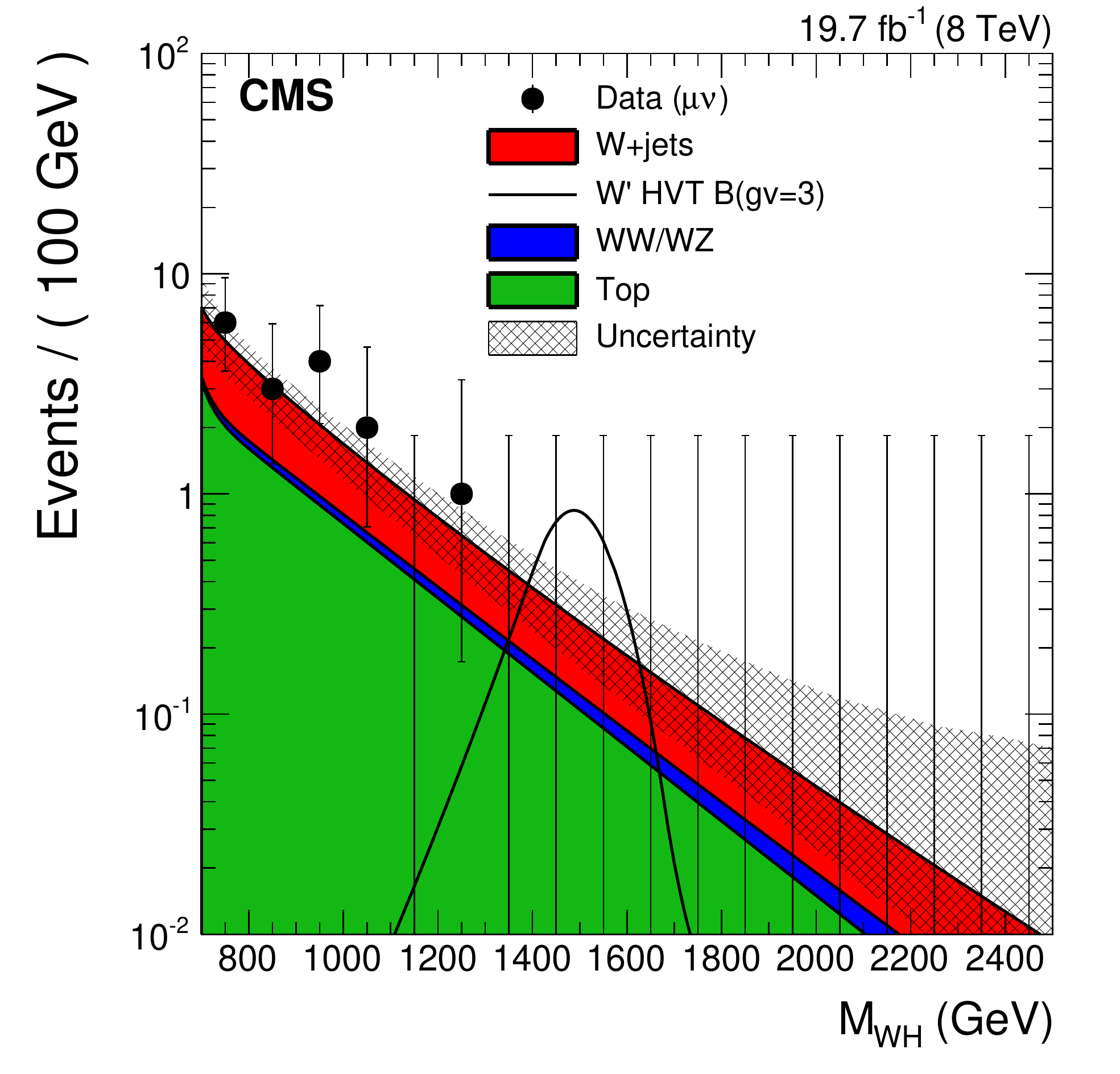}
\includegraphics[width=0.45\linewidth]{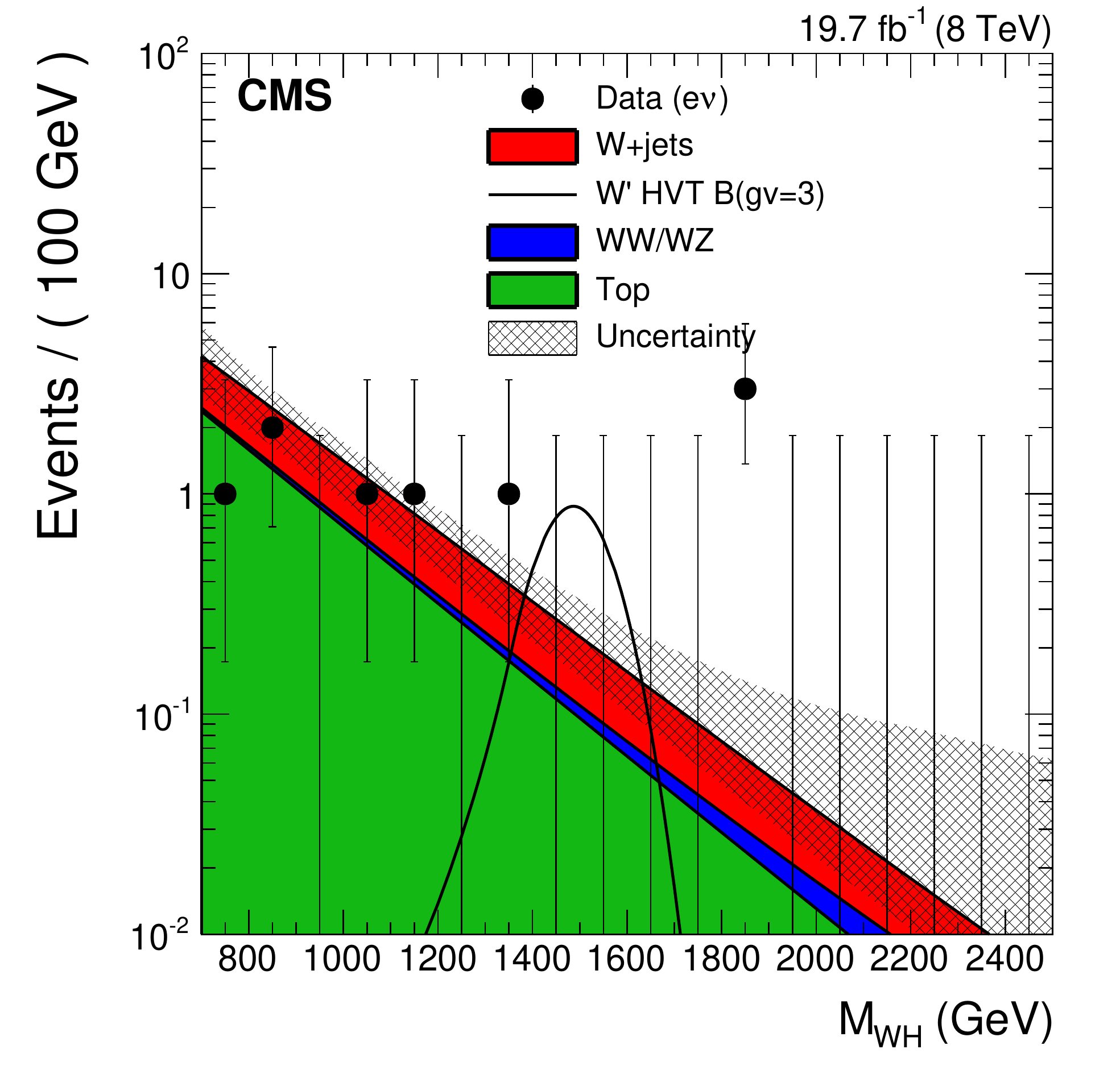}
\caption{
Final distributions in \mWH for data and expected backgrounds for both the muon
(left) and the electron (right) categories.
}
\label{fig:MZZwithBackgroundWH}
\end{figure}

We search for resonances on top of a smoothly falling background distribution mainly composed of \ttbar and W+jets events.
For the resonance mass range considered, the two quarks from the Higgs boson decay are separated by a small angle
in space, resulting in the presence of one single merged jet after hadronization.
This jet is tagged as coming from a Higgs boson
through the study of its mass, applying jet substructure techniques~\cite{JME-13-006}, and
dedicated b-tagging techniques for high transverse momentum Higgs bosons~\cite{CMS:BTV13001}.
This is the first search for high mass WH resonances in the semi-leptonic final state and
one of the first searches deploying jet substructure techniques and dedicated b-tagging techniques
for high transverse momentum Higgs bosons.
After reconstructing the W and Higgs bosons, we apply the final
selections used for the search. Both the W and the Higgs boson
candidates must have a \pt greater than 200~GeV.
Figure~\ref{fig:MZZwithBackgroundWH} shows the final observed spectrum
in \mWH of the selected events in the two lepton categories.
The highest-mass data event is from the electron category and it has $m_{\PW\PH}\approx 1.9$~TeV.
The observed data and the predicted background in the muon channel agree with each other.
In the electron channel an excess of 3 events are observed with $m_{\PW\PH} > 1.8$~TeV,
where less than 0.3 events are expected, while in the muon channel no events with $m_{\PW\PH} > 1.8$~TeV are observed.
The measured pseudo-rapidity values of the jet in the 3 electron channel
events with highest \mWH\ are 0.44, 0.84, 1.87,
while for a W$^\prime$ resonance less than 2\% of the events are
expected to have a pseudo-rapidity above 1.8.
The significance of this excess is discussed after the description of the systematic uncertainties.

\begin{figure}[htbp]
\centering
\includegraphics[width=0.45\linewidth]{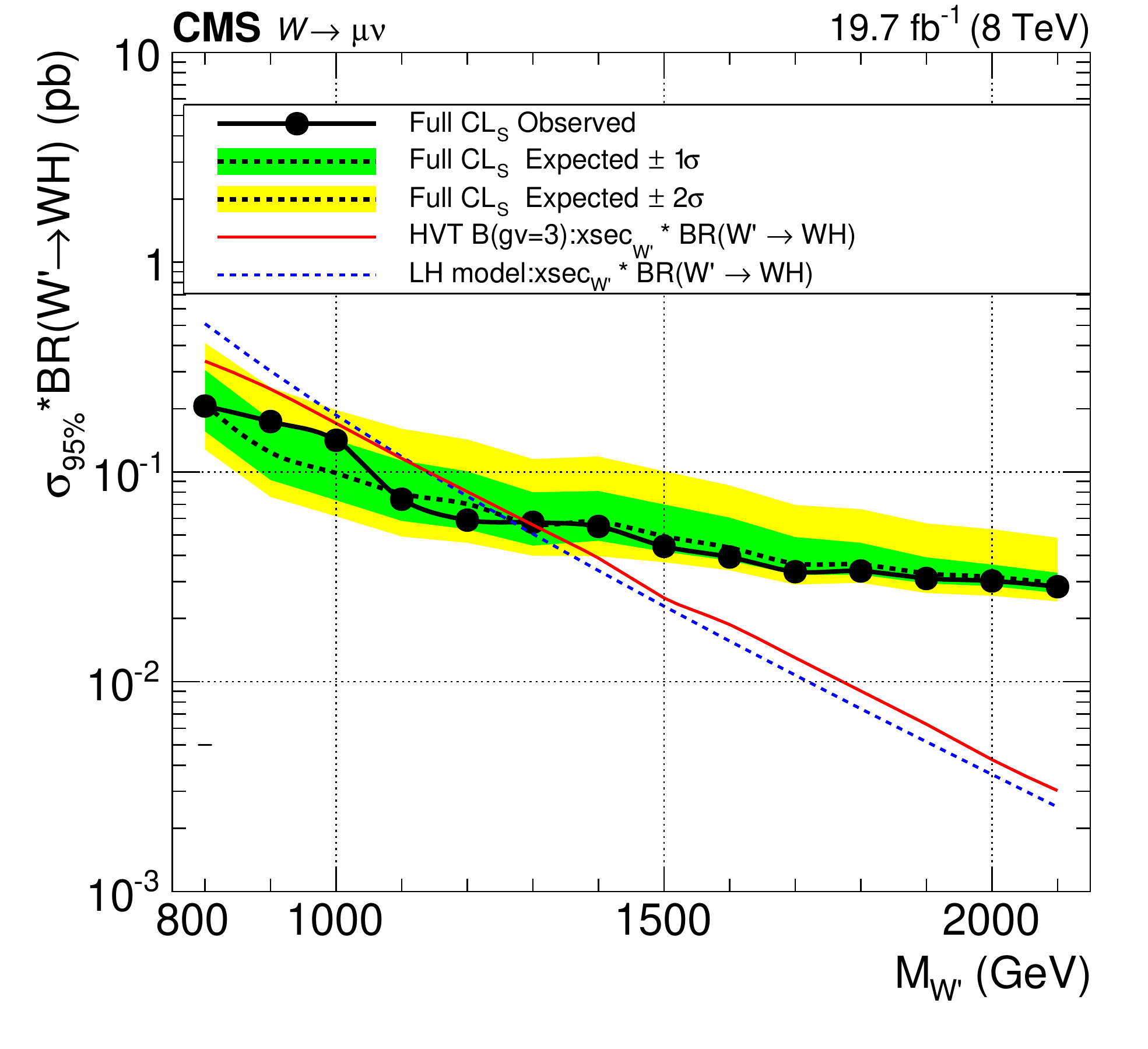}
\includegraphics[width=0.45\linewidth]{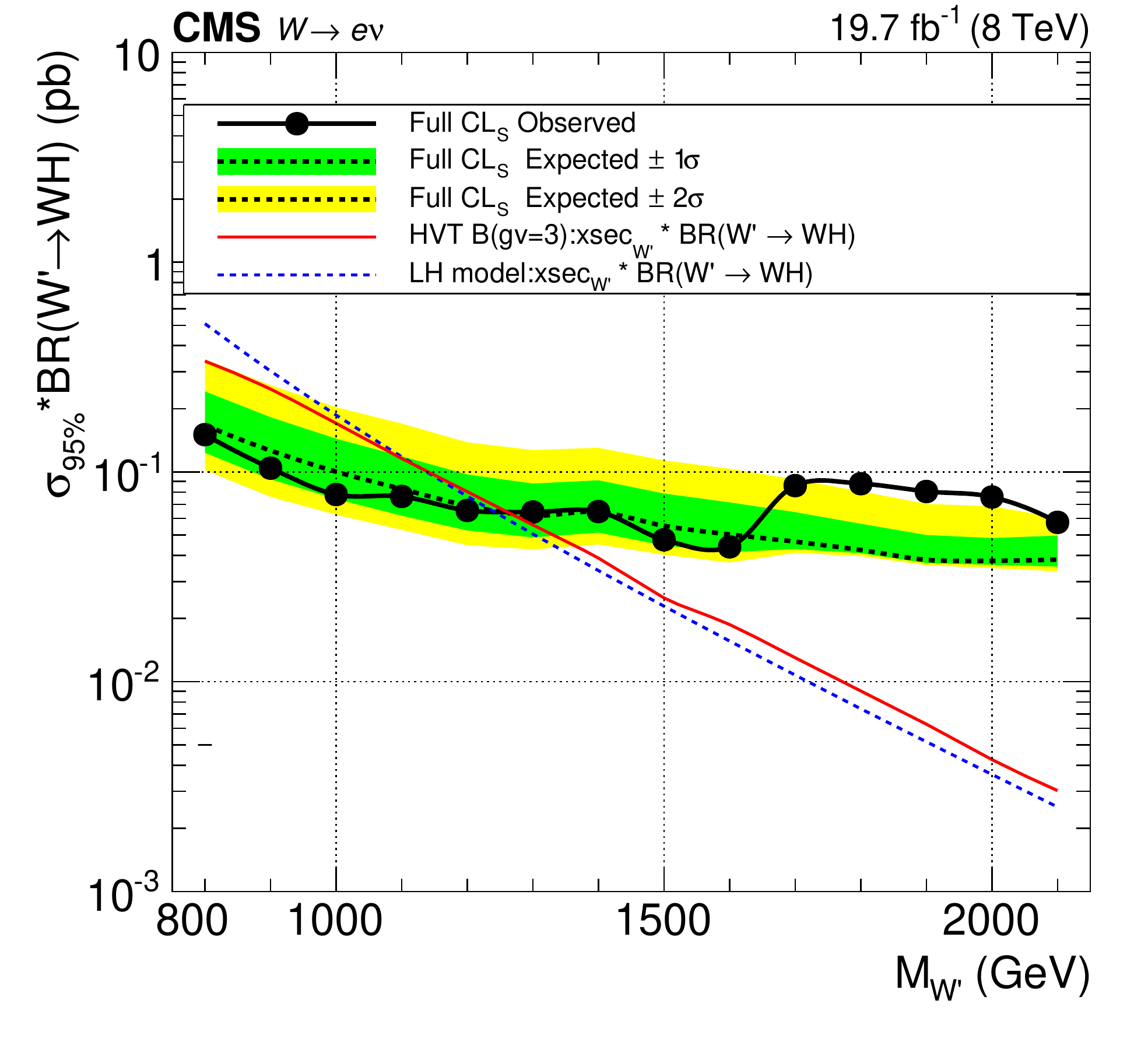}
\caption{
  Observed (solid) and expected (dashed) 95\% CL upper limits (CLs) on the
  product of the W' production cross section and the branching
  fraction of $\PWpr\to \PW\PH$ for muon (left) and electron (right) channels.
  The cross section for the production of a $\PWpr$ in the Little Higgs model and the HVT scenario B
  multiplied by its branching fraction for the relevant process is overlaid.
}
\label{fig:seperateLimits_FullCLs}
\end{figure}

We set upper limits on the production cross section
using \CLs. 
Exclusion limits can be set in the context of the $\PWpr$ model,
under the assumption of a natural width negligible with respect to the
experimental resolution (narrow-width approximation).
Figure~\ref{fig:seperateLimits_FullCLs} shows the 95\% CL expected and observed
exclusion limits for the electron and muon channels separately.
The limits are compared with the
cross section times the branching fraction to $\Wo\Ho$  for
a $\PWpr$ from the Little Higgs model and the HVT scenario B.
The observed (expected) lower limit on the $\PWpr$ mass is 1.4 (1.4)~TeV in the Little Higgs model
and 1.5 (1.5)~TeV in the HVT scenario B.
The full study can be found in Ref.~\cite{CMS:2015gla}.

\section{Search for Narrow Resonances using the Dijet Mass Spectrum}

Many extensions of the SM predict the existence of new massive objects that
couple to quarks or antiquarks (q) and gluons (g), thus resulting in
resonances in the dijet mass spectrum. 
The results presented in this section extend the search to higher values of the
resonance masses. We consider the following specific models of narrow $s$-channel dijet
resonances: string resonances
(S)~\cite{Anchordoqui:2008di,Cullen:2000ef}; scalar diquarks
(D)~\cite{ref_diquark}; excited quarks
(\Qstar)~\cite{ref_qstar,Baur:1989kv}; axigluons
(A)~\cite{ref_axi,Chivukula:2011ng}; color-octet colorons
(C)~\cite{ref_coloron}; the s8 resonance predicted in
technicolor~\cite{Han:2010rf}; new gauge bosons ($\mbox{W}^\prime$ and
$\mbox{Z}^\prime$)~\cite{ref_gauge}; Randall-Sundrum (RS) gravitons
(G)~\cite{ref_rsg}. More details on the specific choices of couplings
for the models considered can be found in Ref.~\cite{CMS:2012yf}.

\begin{figure}[hbt]
  \centering
    \includegraphics[width=0.55\linewidth]{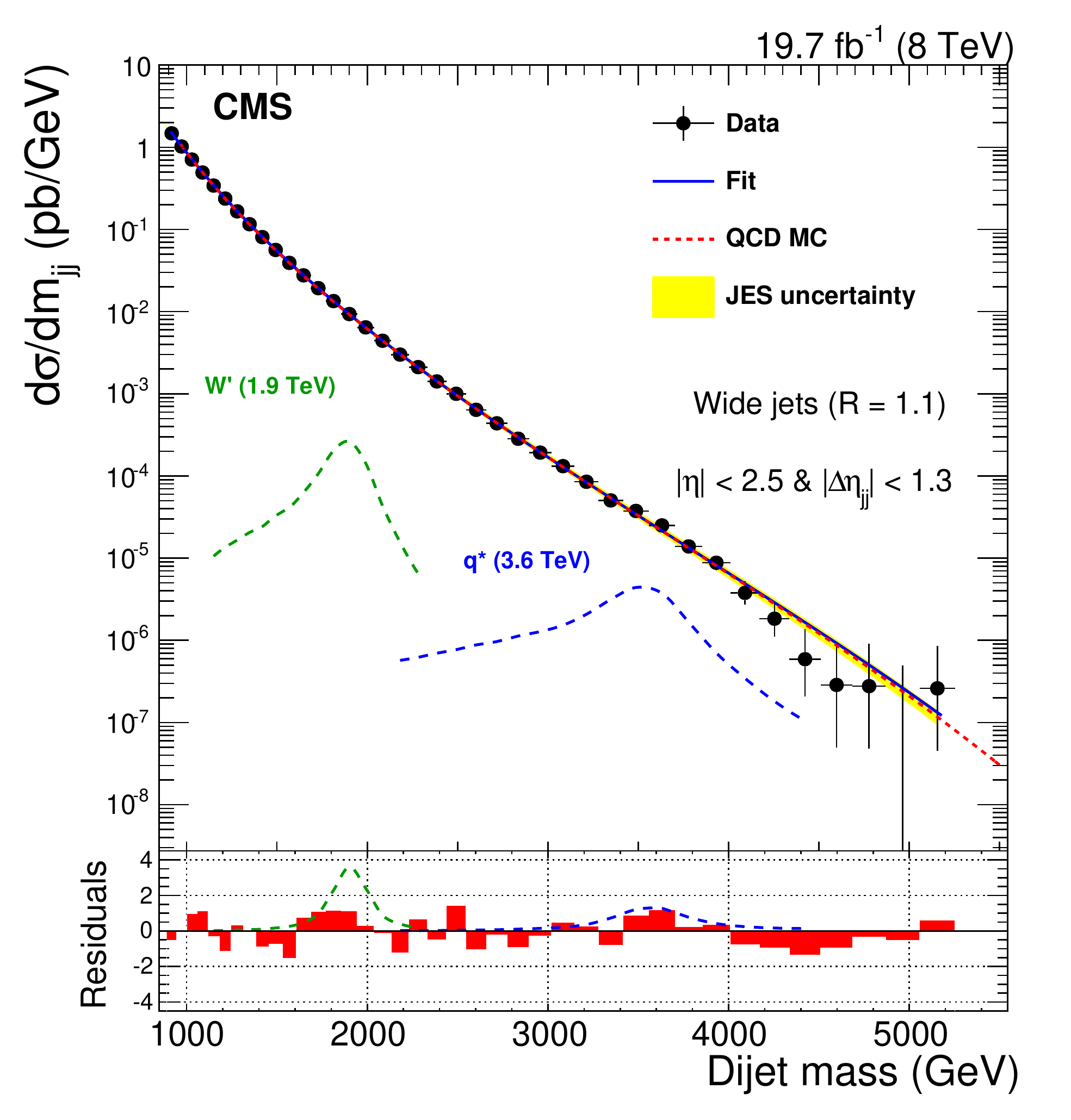}
    \caption{Inclusive dijet mass spectrum from wide jets
      (points) compared to a fit (solid curve) and to
      predictions including detector simulation of
      multijet events and signal resonances.  The predicted multijet
      shape (QCD MC) has been scaled to the data (see text).
      The vertical error bars are statistical only
      and the horizontal error bars are the bin widths.
      For comparison,the signal distributions for a \PWpr resonance of
      mass 1900\GeV and an excited quark of mass 3.6\TeV are shown.
      The bin-by-bin fit residuals scaled to the statistical uncertainty of the data,
      $(\text{data}-\text{fit})/\sigma_{\text{data}}$, are shown at the
      bottom and compared with the expected signal contributions.
    }
      \label{figDataAndMC}
  \end{figure}

A method based on data is used to estimate the background from multijet
production. We fit the following parameterization to the data:
\begin{equation}
\frac{\rd\sigma}{\rd\mjj}=
\frac{P_{0} (1 - x)^{P_{1}}}{x^{P_{2} + P_{3} \ln{(x)}}}
\label{eq:bkgfunction}
\end{equation}
with the variable $x=\mjj/\sqrt{s}$ and four free parameters
$P_0$, $P_1$, $P_2$, and $P_3$.  
A Fisher $F$-test is used to confirm that no
additional parameters are needed to model these distributions for a
data set as large as the available one.  The fit of the data to the
function given in Eq.~(\ref{eq:bkgfunction}) returns a chi-squared
value of 30.65 for 35 degrees of freedom. The difference
between the data and the fit value is also shown at the bottom of
Fig.~\ref{figDataAndMC}, normalized to the statistical uncertainty of
the data. The function of
Eq.~(\ref{eq:bkgfunction}) is also fit to these data distributions.
The data are well described by this function and no significant
deviations from the background hypothesis are observed.

\begin{figure*}[hbtp]
  \centering
    \includegraphics[width=0.55\linewidth]{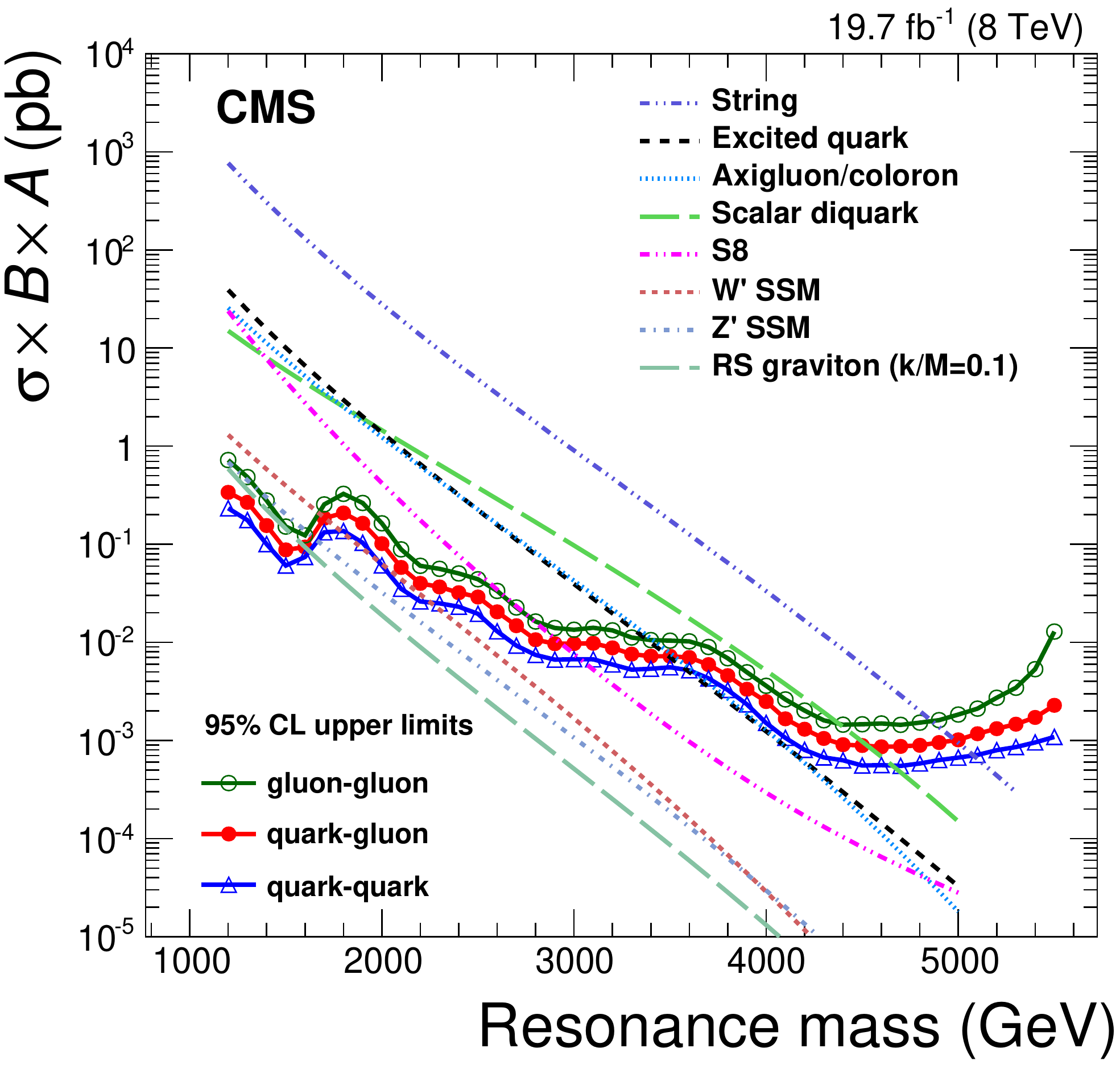}
    \caption{The observed 95\% CL upper limit
      for narrow dijet resonances. Top: limit on gluon-gluon,
      quark-gluon, and quark-quark narrow resonances from the inclusive
      analysis, compared to LO theoretical predictions for string
      resonances~\cite{Anchordoqui:2008di,Cullen:2000ef},
      excited quarks~\cite{ref_qstar,Baur:1989kv},
      axigluons~\cite{ref_axi,Chivukula:2011ng},
      colorons~\cite{ref_coloron},
      scalar diquarks~\cite{ref_diquark},
      S8 resonances~\cite{Han:2010rf},
	  gravitons.
	  Bottom left: combined limits on gg/bb resonances. The
      theoretical cross section for an RS graviton is
      shown for comparison.  Bottom right: combined limits on qq/bb resonances
      \label{figExpectedLimit}}
  \end{figure*}

The expected limits on the cross section are estimated with
pseudo-experiments generated using background shapes, which are
obtained by signal-plus-background fits to the data.
Figure~\ref{figExpectedLimit} shows the expected limits and their
uncertainty bands. 
The full study can be found in Ref.~\cite{Khachatryan:2015sja}.

\section{Search for dark matter in the V/jet + \ETm final state}

Dark matter (DM) is one of the most compelling pieces of indirect evidence for BSM physics.
In many DM theories, the collider production of particle DM proceeds via a mediator with couplings to the SM.
The canonical ``monojet'' search strategy provides a model-independent means of exploring this scenario~\cite{monojet1,monojet2} while
related ``mono-V'' (V=W/Z) searches~\cite{monolep,monoZHbb,Aad:2014vka,Aad:2013oja,ATLAS:2014wra} target the associated production of DM with SM vector bosons,
which can be enhanced in theories with non-universal DM couplings~\cite{IVDM}.  The interpretation of results from these and other DM searches at
the Large Hadron Collider (LHC) have generally utilized effective field theories (EFT)
that assume heavy mediators and DM production via contact interactions.

The analysis described is a search for new physics containing an energetic jet and \ETm using the full 8 TeV dataset from Run-1 corresponding
to an integrated luminosity of 19.7\fbinv.
The search is the first at CMS to target the hadronic decay modes of the vector bosons in the mono-V channels.
A multivariate V-tagging technique is employed to identify the individually resolved decay products of moderately boosted vector bosons.
The exploration of mono-V production at high boost utilises recently developed techniques designed to exploit information available in the sub-structure
of jets.
The analysis incorporates both monojet and mono-V final states in a combined search, categorized according to the nature of the jets in the event. 
The DM signal extraction is performed by considering the shape of the $\ETm$ distribution in each event category, which provides improved sensitivity compared to the previous
CMS monojet analysis~\cite{monojet1}.

The results are interpreted using simplified DM models, which are most appropriate when considering DM mediator production~\cite{Buchmueller:2013dya},
which span a broad range of mediator and DM particle properties. Where previous searches have provided interpretations using an EFT based approach, in light of recent developments,
this analysis focuses on the use of the simplified models scheme for which a mediator and coupling are well defined~\cite{Abercrombie:2015wmb}.

Signal events are selected on the basis of large values of \ETm and one or more high-\pt jet(s),
explicitly vetoing isolated leptons and photons.
Selected events are classified into three event categories, based on the topology of the jets, in order to distinguish between
initial or final state radiation of gluons or quarks versus jets arising from hadronic vector boson decays.
Events are first scrutinized for the presence of an unresolved vector boson, subsequently for a resolved vector boson
and finally the remaining events are collected into the monojet category.

If the electroweak boson decays hadronically and it has sufficiently high transverse momentum, both its hadronic
decay products are captured by a single reconstructed ``fat''-jet.
The selection for this category is devised to identify such events by selecting events containing a fat-jet with a large $\pt$.
Events which contain additional jets close to the fat-jet, but no closer than $\Delta R < 0.5$,
are selected to include the frequent cases in which initial state radiation yields additional jets.
In cases where the electroweak boson has insufficient transverse momentum for its hadronic
decay to be fully contained in a single reconstructed fat-jet, a selection that looks for decays
into a pair of jets is applied to recover the event.
The selection requires that each jet has $p_T>30$ GeV and $|\eta|<2.5$ and that the dijet has a mass between 60\gev and 110\gev, consistent with originating
from an electroweak boson.
The events that do not qualify for either of the two V-tagged categories are tested for the presence of a single jet originating from quark or gluon radiation.
For the monojet category, at least one jet within $|\eta|<2.0$ with \pt greater
than 150 \gev and a \ETm greater than 200 \gev is required.

\begin{figure}[htbp]
  \centering
    \includegraphics[width=0.45\linewidth]{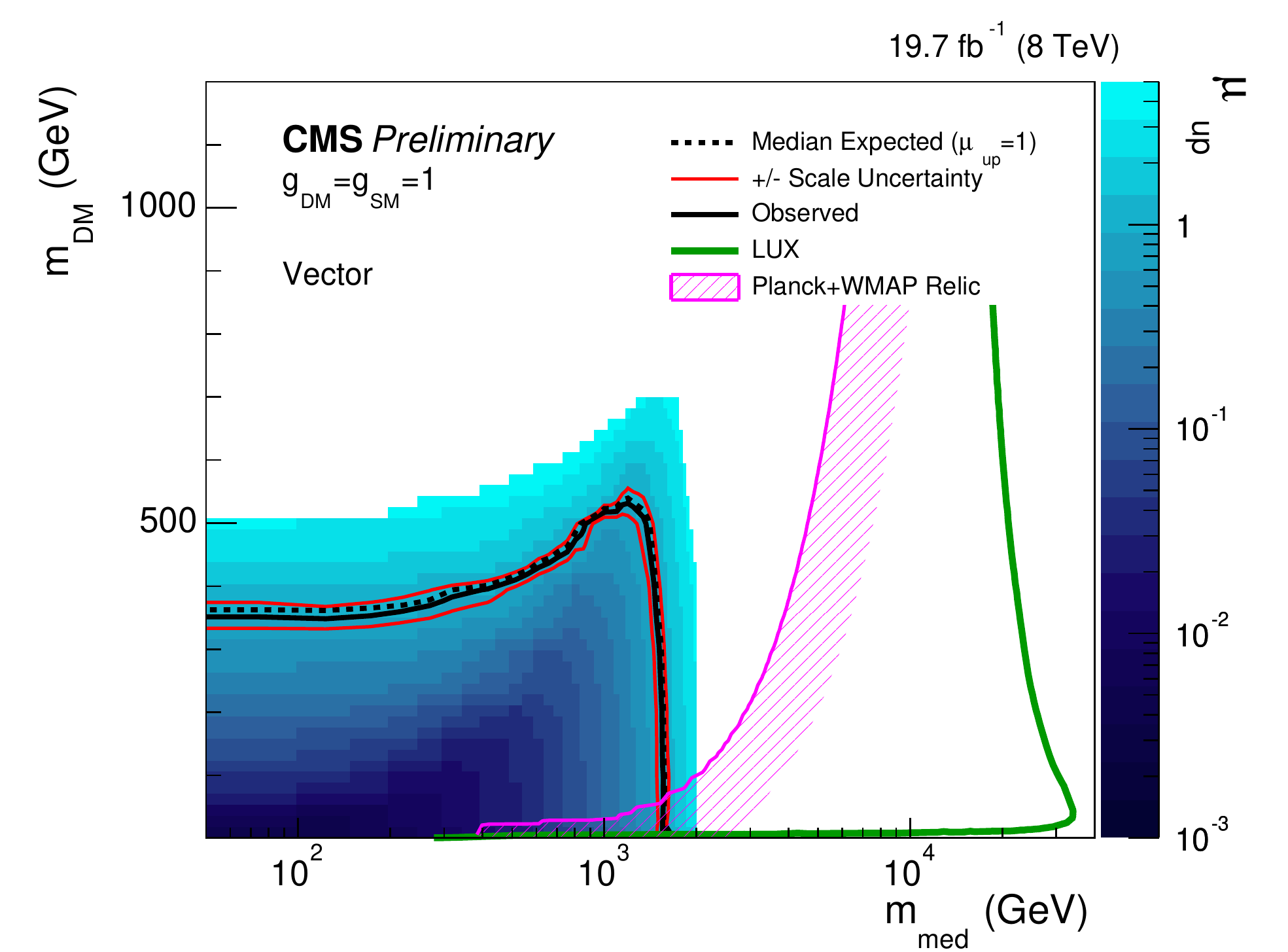}
    \includegraphics[width=0.45\linewidth]{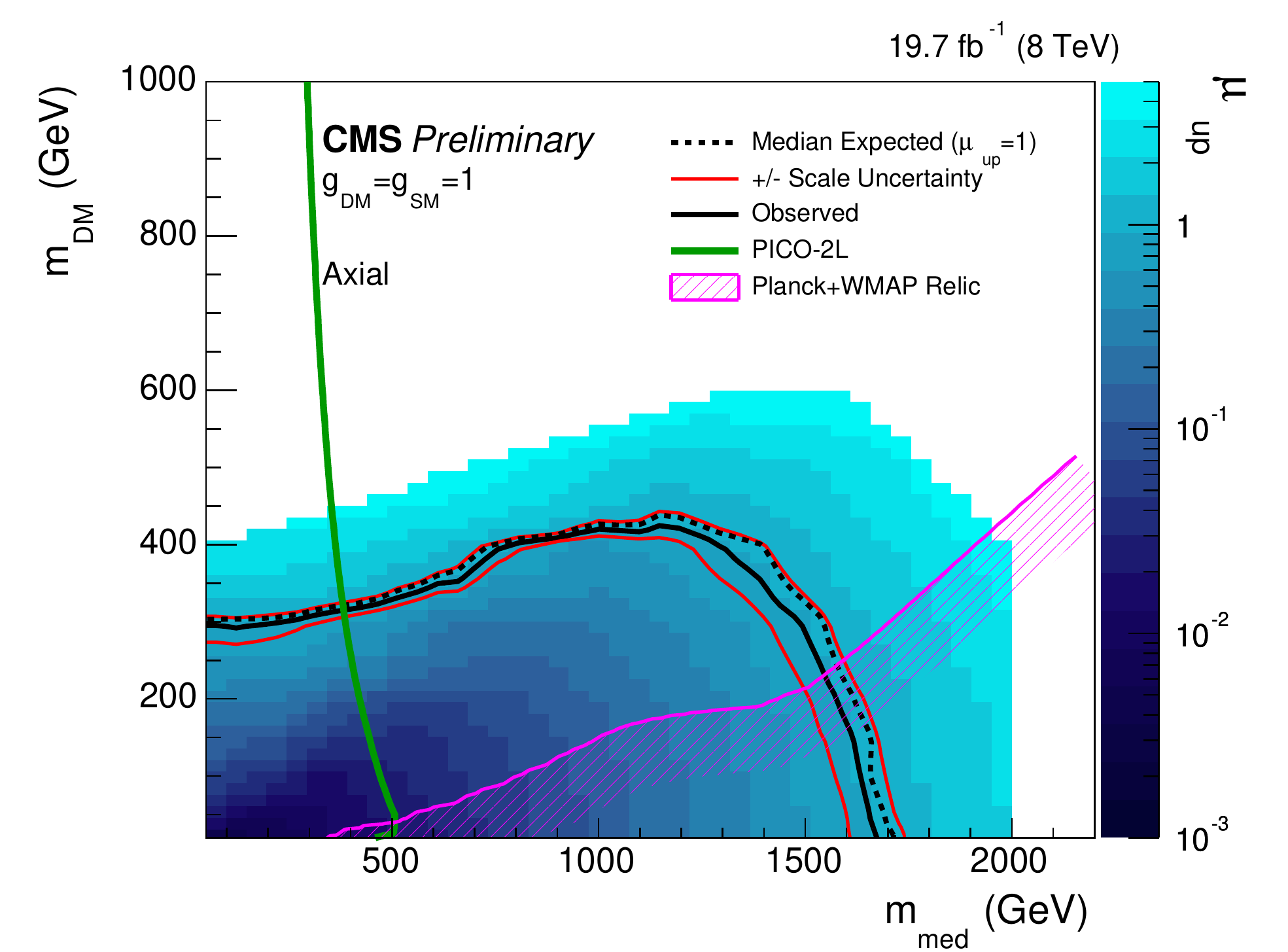}
    \caption{90\% CL Exclusion contours in the $m_{\textrm{med}}-m_{\textrm{DM}}$ plane assuming a vector (a), axial-vector (b) mediator.
The blue scale shows the 90\% CL upper limit on the signal strength assuming the mediator only couples to fermions. For the scalar and pseudoscalar mediators, the exclusion
contour assuming coupling only to fermions is explicitly shown in the orange line. The white region shows model points which were not tested when assuming coupling only
to fermions and are not expected to be excluded by this analysis under this assumption.
The excluded region is to the bottom-left of the contours shown in all cases except for that from the relic density as indicated by the shading.
In all of the mediator models, a minimum width is assumed.
\label{fig:masslims}}
\end{figure}

Agreement between the expected SM backgrounds and data is observed at the percent level across the three categories.
The local significance is calculated by constructing a signal model for which the signal events exist only in a single bin and comparing the
likelihood of the background-only fit to a fit including that signal, allowing the signal yield to float freely.
The largest local significance, calculated this way, in any of the bins across the three categories, is 1.9$\sigma$ and corresponds to the excess seen in the last \ETm bin of the
monojet category.
Limits are calculated in terms of exclusion of regions in the $m_{\mathrm{MED}}-m_{\mathrm{DM}}$ plane, assuming the four different mediators,
by determining the points for which $\mu\ge1$ is excluded at 90\% CL or more. Figure~\ref{fig:masslims} shows the 90\% CL exclusions for the vector, axial-vector,
scalar and pseudoscalar mediator models.  The 90\% CL upper limit on the ratio of excluded cross-section to the predicted cross-section ($\mu_{up}$), when assuming the mediator only
couples to fermions, is shown by the blue color scale. These limits are obtained under the assumption that only the initial state partons and the DM
particle contribute to the width of the mediator. For all models, the width is fixed under the minimum width
constraint~\cite{An:2012va,Abercrombie:2015wmb}. For the vector
mediator, the direct detection bounds dominate across most of the plane, while for the axial-vector, there is good complementarity between the direct detection limits and
those from this analysis. Limits in the scalar mediator scenario are much weaker than those from direct detection for small dark-matter masses. Additional sensitivity is gained
for larger mediator masses in this scenario above 350 \GeV due to the rise in cross-section of the gluon fusion loop process above the $t\bar{t}$ threshold.
In the pseudoscalar mediator scenario, the limits from this analysis exceed the reach in $m_{MED}$ than those from FermiLAT.
The full study can be found in Ref.~\cite{CMS:2015jha}.

\section{Summary} 

CMS has carried out several searches for BSM physics with an
integrated luminosity of around 20 fb$^{-1}$ taken at $\sqrt{s}=8~$\TeV. 
No significant excess in data with respect to the SM expectation has been observed so far. 
However, searches in large regions of the parameter space are still
in progress. Subsequently, new results obtained at $\sqrt{s}=13~$\TeV\ will be looked upon with eager anticipation.

\bigskip 

\bibliography{}

\end{document}